\documentclass[12pt, a4paper, oneside]{article}
\usepackage{cite}
\usepackage{graphicx}
\usepackage{epsfig}
\usepackage{graphics}
\usepackage{epstopdf}
\usepackage{amssymb,amsmath}
\usepackage[a4paper=true,pagebackref=false]{hyperref}
\usepackage[scale=0.8]{geometry}
\usepackage[labelfont=bf,textfont=md]{caption}
\hypersetup{colorlinks = true, linkcolor = magenta, anchorcolor = red, citecolor = cyan, filecolor = red, pagecolor = red, urlcolor = red}

\title{Dynamical Systems Analysis of  $f(R, L_m)$ Gravity Model}

\author{Aman Shukla\footnote{amanshukla@bhu.ac.in},    R Chaubey\footnote{rchaubey@bhu.ac.in}\\
\\
Centre for Interdisciplinary Mathematical Sciences \\ Institute of Science \\ Banaras Hindu University\\ Varanasi, India. 
\\
\\
Rakesh Raushan\footnote{rakesh.raushan4@bhu.ac.in}
\\
Department of Mathematics, BMA College\\ LNM University, Darbhanga, Bihar.
}

\date{}

\begin{document}
\maketitle

\begin{abstract}

In this article, we examine the dynamical evolution of flat FRW cosmological model in  $f(R, L_m)$ gravity theory. We consider the general form of $f(R, L_m)$ defined as $f(R, L_m) = \Lambda + \frac{\alpha}{2} R + \beta L_m^n$, where $\Lambda$, $\alpha$, $\beta$, $n$ are model parameters, with the matter Lagrangian given by $L_m = -p$.  We investigate the model through phase plane analysis, actively studying the evolution of cosmological solutions using dynamical systems techniques. To analyse the evolution equations, we introduce suitable transformations of variables and discuss the corresponding solutions by phase-plane analysis. The nature of critical points is analysed and stable attractors are examined for $f(R, L_m)$ gravity cosmological model. We examine the linear and classical stability of the model and discuss it in detail. Further, we investigate the transition stage of the Universe, i.e. from the early decelerating stage to the present accelerating phase of the Universe by evolution of the effective equation of state, $\{r,s\}$ parameters and statefinder diagnostics for the central values of parameters $\alpha, \beta$ and $n$ constrained using MCMC technique with cosmic chronometer data.

\end{abstract}



\noindent\hrulefill
\section{Introduction}
\label{sec1}
According to astronomical observations, there are evidences indicating that the Universe is undergoing accelerated expansion \cite{AR, P, NA}. The standard model is based on the assumption that the gravitational mass of the Universe is positive. The accelerated expansion of the Universe can potentially be explained by modifying the standard cosmological model \cite{CDL, BCN, NOO, BBC, sajal2}. The current standard model, known as the $\Lambda$ Cold Dark Matter ($\Lambda$CDM) model, incorporates pressure-less cold dark matter and dark energy, which is attributed to a positive cosmological constant. This model accounts for various characteristics observed in the Universe. In this context, dark energy refers to a component of the cosmological model that exerts negative pressure. In the context of explaining late-time acceleration, this approach brings attention to several intriguing aspects of the cosmological modelling of the universe's evolution \cite{CDCT, SDE, GLD, RA, CAAT1, CAAT2, BFFG, SRC, OO, LPD, ASINGH, LWU, RPM, MAC, FB2023, Cap2022}.

A compelling approach to explain the recent findings regarding the expansion of the Universe is to propose that the conventional models of Einstein's General relativity become inadequate when applied to vast cosmic scales. Instead, a broader framework is needed to describe the gravitational field. Numerous methods exist for extending the Einstein-Hilbert action of General relativity. Theoretical models have been introduced, such as replacing the standard action with a more comprehensive function $f(R)$, where $R$ represents the Ricci scalar, discussed in several literature \cite{BA, KG}. The scenario of cosmic expansion at late times can be effectively described using $f(R) $ gravity \cite{SC}. The limitations and requirements for viable cosmological models have been thoroughly examined  \cite{SC1, APT}. It has been demonstrated that $f(R) $ gravity models that satisfy the constraints of solar system tests do indeed exist \cite{NO, F, Z, AT}. In addition, authors \cite{TS, LZZ, CT, C, SJ} present observational manifestations of $f(R)$ dark energy models, as well as the constraints imposed by the solar system and the equivalence principle on $f(R)$ gravity. Moreover, there have been discussions \cite{NO1, NO2, GC} about other $f(R)$ models that unify early inflation with dark energy and successfully pass local tests. For exploring various cosmological implications of the $f(R)$ gravity model, some literature \cite{JS, CCS, RN}  can be consulted.

A proposal was made for a modification of the $f(R)$  gravity theory, which incorporates a direct connection between the matter Lagrangian density $L_m$ and a generic function $f(R)$ \cite{OB}. This coupling between matter and geometry gives rise to an additional force perpendicular to the four-velocity vector when massive particles exhibit non-geodesic motion. The model was further extended to encompass arbitrary couplings in both matter and geometry \cite{H1}. Extensive investigations into the cosmological and astrophysical implications of these non-minimal matter-geometry couplings have been conducted  \cite{H2, H3, SN, F1, F2}. Harko and Lobo \cite{H4} introduced a more advanced generalization of matter-curvature coupling theories known as the $f(R, L_m)$ gravity theory, where $f(R, L_m)$ represents an arbitrary function of the matter Lagrangian density $L_m$ and the Ricci scalar $R$. The $f(R, L_m)$ gravity theory represents the most extensive expansion of gravitational theories formulated in Riemann space. In this theory, the motion of test particles deviates from the geodesic path, resulting in the emergence of an additional force perpendicular to the four-velocity vector. The $f(R, L_m)$ gravity models exhibit a notable violation of the equivalence principle, which has been rigorously tested within the solar system \cite{F3, BPT}. The study by Wang and Liao has explored the energy conditions within the framework of $f(R, L_m)$ gravity \cite{WL}. Additionally, Gonclaves and Moraes \cite{GM} investigated cosmology considering the non-minimal coupling between matter and geometry, incorporating the $f(R, L_m)$ gravity theory.  In the present study, we will consider the general form of $f(R, L_m)$  with $L_m = -p$ and analyze the qualitative behavior of the model.

This present paper is structured into the following five sections. Section $\ref{sec1}$ serves as an introduction, providing an overview of the research topic. Section $\ref{sec2}$ focuses on the formulation of the $f(R, L_m)$ gravity, delving into its theoretical framework and principles. Section $\ref{sec3}$ focuses on utilizing the perfect-fluid stress-energy-momentum tensor to derive the field equations applicable to a flat FLRW universe.  In this section, we obtained the cosmological solution for the $f(R, L_m) = \Lambda + \frac{\alpha}{2} R + \beta L_m^n$ model. Section $\ref{sec4}$ is dedicated to formulation of autonomous equations of the system and  exploring the dynamical system constraints. We also provide an analysis of the critical points and phase plane behavior of the model. In this section, we also employ emcee Python package and find the corresponding posterior phase space of parameters. The results of dynamical systems approach are used to explore the observational viability of the model with related constraints on physical parameters. Conclusions are given in the last Section $\ref{sec5}$.

\section{\textbf{$f(R,L_m)$  Theory of Gravity}}
\label{sec2}
The action which defines the gravitational interactions in $f(R,L_m)$ gravity is given as
\begin{eqnarray}
    I = \int f(R,L_m) \sqrt{-g} d^4 x
    \label{eq1}
\end{eqnarray}
where $L_m$ is the matter Lagrangian, $R$ is Ricci scalar and $f(R,L_m)$ represents the arbitrary function of $R$ and $L_m$.

The Ricci scalar $R$ defined as 
\begin{eqnarray}
    R =  g_{}^{ij} R_{ij}
    \label{eq2}
\end{eqnarray}
where $g_{}^{ij}$ is the contravariant form of the metric tensor and $R_{ij}$ represents Ricci tensor which can be described in the following manner 
\begin{eqnarray}
    R_{ij} = \frac{\partial^2}{\partial x^i\partial x^j} \ln\sqrt{-g} - \frac{\partial\Gamma_{ij}^{k}}{\partial x^k} +\Gamma_{ik}^{l} \Gamma_{jl}^{k} - \Gamma_{ij}^{k} \frac{\partial}{\partial x^k} \ln \sqrt{-g}
    \label{eq3}
\end{eqnarray}
where $\Gamma_{ij}^{l}$ represents the well known Levi-Civita connection defined as

\begin{eqnarray}
    \Gamma_{ij}^{l}= \frac{g^{lk}}{2} \left[\frac{\partial g_{ik}}{\partial x^j}+ \frac{\partial g_{jk}}{\partial x^i}-\frac{\partial g_{ij}}{\partial x^k}\right]
    \label{eq4}
\end{eqnarray}

After varying the action Eq.(\ref{eq1}) over the metric tensor $g_{ij}$, we acquire the following field equation 
\begin{eqnarray}
    f_R R_{ij} + (g_{ij} \nabla_{k}  \nabla_{}^{k} - \nabla_{i}  \nabla_{j})f_R-\frac{1}{2}(f-f_{L_{m}} L_m) g_{ij} = \frac{1}{2} f_{L_{m}}T_{ij}
    \label{eq5}
\end{eqnarray}
where $f_R = \frac{\partial f}{\partial R}$, $f_{L_m} = \frac{\partial f}{\partial f_{L_m}}$ and $T_{ij}$  represents the energy-momentum tensor for the perfect type fluid which is expressed as 

\begin{eqnarray}
    T_{ij} = \frac{-2}{\sqrt{-g}} \frac{\delta(\sqrt{-g} L_m)}{\delta g^{ij}}
    \label{eq6}
\end{eqnarray}
We can establish a relationship between energy momentum tensor $T_{ij}$, Ricci scalar $R$, and matter Lagrangian density $L_m$ by using the contraction on the field equation ($\ref{eq5}$).
The required relation is exhibited as
\begin{eqnarray}
    R f_R + 3 \square f_R - 2(f-f_{L_m} L_m)= \frac{1}{2} f_{L_m} T_{ij}
    \label{eq7}
\end{eqnarray}
where $\square F = \frac{1}{\sqrt{-g}} \partial_l(\sqrt{-g}g^{lj} \partial_i F)$ for any arbitrary scalar function $F$. 

The Covariant derivative of equation ($\ref{eq5}$) leads to
\begin{eqnarray}
    \nabla^i T_{ij} = 2 \nabla^i \ln(f_{L_m}) \frac{\partial L_m}{\partial g^{ij}}
    \label{eq8}
\end{eqnarray}
where $\frac{\partial L_m}{\partial g^{ij}} = -\frac{1}{2}(g_{ij} L_m - T_{ij})$

\section{Model and Basic Equations}
\label{sec3}
We consider the following flat Friedman-Lamatre-Robertson-Walker (FLRW) metric \cite{R} for the homogeneous spatial universe as

\begin{eqnarray}
    ds^2 = -dt^2 +a^2(t)\left[dx^2+dy^2+dz^2\right]
    \label{eq9}
\end{eqnarray}
where $a(t) $ is the scale factor quantifying the extent of cosmic expansion at a specific time $t$.

The non-vanishing components of the Christoffel symbols correspond to the metric ($\ref{eq9}$)
\begin{eqnarray}
    \Gamma^0_{ij} = -\frac{1}{2} g^{00} \frac{\partial g_{ij}}{\partial x^0}, 
    \quad \Gamma^k_{0j} = \Gamma^k_{j0} = \frac{1}{2} g^{k\lambda} \frac{\partial g_{j\lambda}}{\partial x^0}
    \label{eq10}
\end{eqnarray}
where $i,j,k = 1,2,3 $

By utilizing equation ($\ref{eq3}$), we can determine the following non-zero components of the Ricci curvature tensor.
\begin{eqnarray}
    R_{0}^0 = 3 \frac{\ddot{a}}{a}, \quad R_{1}^1 = R_{2}^2 = R_{3}^3 = \frac{\ddot{a}}{a} + 2 \left(\frac{\dot{a}}{a}\right)^2
    \label{eq11}
\end{eqnarray}
Hence, the Ricci scalar $R$ associated with the line element ($\ref{eq9}$) can be derived as follows

\begin{eqnarray}
    R = 6 \frac{\ddot{a}}{a} + 6 \left(\frac{\dot{a}}{a}\right)^2 = 6(H\dot{H} + 2H^2)
    \label{eq12}
\end{eqnarray}
where $H = \frac{\dot{a}}{a}$ is the Hubble parameter.

The energy-momentum tensor that describes the matter content of the Universe, which is filled with a perfect fluid, for the given line element ($\ref{eq9}$), can be expressed as follows
\begin{eqnarray}
    T_{ij} = (\rho + p)u_iu_j + pg_{ij}
    \label{eq13}
\end{eqnarray}
where $p$ represents the pressure exerted by the cosmic fluid, $\rho$ corresponds to the energy density, $g_{ij}$ refers to the metric tensor, and $u^i = (1,0,0,0)$ represents the components of the co-moving four-velocity vector in the cosmic fluid, with $u_iu^i$ equal to $ -1$.

The equations that govern the dynamics of the Universe in $f(R, L_m)$ gravity\cite{H5}, known as the modified Friedmann equations, can be expressed as follows:

\begin{eqnarray}
    R_{0}^0 f_R - \frac{1}{2}(f - f_{L_m}L_m) + 3H\dot{f_R}= \frac{1}{2}f_{L_m}T_{0}^0 
    \label{eq14}\\
    R_{i}^i f_R - \frac{1}{2}(f - f_{L_m}L_m) + 3H\dot{f_R}= \frac{1}{2}f_{L_m}T_{i}^i; \quad i=1,2,3.
    \label{eq15}
\end{eqnarray}

Here, we consider the gravitational field can be described by a Lagrangian density of the form
\begin{eqnarray}
    f(R, L_m) = \Lambda \exp  \left(\frac{\alpha}{2\Lambda} R + \frac{\beta}{\Lambda} L_m^n\right)
    \label{eq1a}
\end{eqnarray}

where $\Lambda>0$ and $\alpha$, $\beta$ are arbitrary non zero constants. For $\alpha=1$, $\beta=1$ \& $n = 1$, the above model (\ref{eq1a}) reduces to simple toy model for $f(R, L_m)$ gravity obtained by Harko \& Lobo \cite{H4}.
In the limit $\frac{\alpha}{2\Lambda} R + \frac{\beta}{\Lambda} L_m^n << 1$, we obtain

\begin{eqnarray}
    f(R, L_m) = \Lambda + \frac{\alpha}{2} R + \beta L_m^n
    \label{eq16}
\end{eqnarray}

From Eqs. ($\ref{eq16}$) when $L_m = -p$ \cite{H6}, the Friedmann equations ($\ref{eq14}$) and ($\ref{eq15}$) governing the dynamics of a matter-dominated universe can be transformed as : 

\begin{eqnarray}
    3\alpha H^2 + \Lambda = \beta \left(n-1 - n\frac{\rho}{p}\right) (-p)^n 
    \label{eq17}\\
    2 \alpha \dot{H} + 3\alpha H^2 + \Lambda = \beta (2n-1)(-p)^n
    \label{eq18}
\end{eqnarray}
Taking the trace of the field equations, one can derive the following matter conservation equation
\begin{eqnarray}
    \dot{\rho} + 3H(\rho + p) = (n-1)(\rho + p)\frac{\dot{p}}{p}
    \label{eq19}
\end{eqnarray}

\section{Qualitative Analysis and Cosmic Dynamics of the Model}
\label{sec4}
This section investigates the universe's dynamic evolution within the model using qualitative techniques. We achieve this by reformulating the cosmological equations into an autonomous system of differential equations. By analyzing the linear stability of critical points associated with specific cosmological solutions, we gain insights into the behaviour and stability of the system.

Using the equation of state parameter $p = \omega \rho$, we reformulate the equation ($\ref{eq17}$) as

\begin{eqnarray}
    1 = - \frac{\Lambda}{3 \alpha H^2} + \frac{\beta}{3 \alpha H^2} \left(n+\omega - n \omega\right) (-\omega)^{n-1} \rho^n
    \label{eq20}
\end{eqnarray}
Using equation ($\ref{eq18}$), we obtain the following expression 
\begin{eqnarray}
    \frac{\dot{H}}{H^2} = \frac{\beta (2n-1)(-\omega)^n \rho^n}{2\alpha H^2} - \frac{\Lambda}{2\alpha H^2} - \frac{3}{2}
    \label{eq21}
\end{eqnarray}
In order to examine the dynamic evolution of the Universe, we introduce the following variables:
\begin{eqnarray}
    x = \frac{\Lambda}{3 H^2}, y = \left(n+\omega - n \omega \right) (-\omega)^{n-1} \frac{\rho^n}{3 H^2}
    \label{eq22}
\end{eqnarray}
From equations ($\ref{eq20}$) and ($\ref{eq22}$), we can obtain the constraint equation expressed in terms of the aforementioned dynamical variables
\begin{eqnarray}
    \alpha + x - \beta y=0
    \label{eq23}
\end{eqnarray}
By considering the constraint equation ($\ref{eq23}$), the resulting state space can be described as having two dimensions. Consequently, we can express the evolution equations as follows:
\begin{eqnarray}
    x' = \frac{dx}{dN} = -3x \left( \frac{\beta (2n-1)\omega}{\alpha (n\omega - \omega -n)} y -\frac{x}{\alpha} -1 \right )
    \label{eq24}
\end{eqnarray}
\begin{eqnarray}
     y' = \frac{dy}{dN} = -3y\left(\frac{\beta (2n-1)\omega}{\alpha (n\omega - \omega -n)} y -\frac{x}{\alpha}-1 + \frac{n(1+\omega)}{1-(n-1)(1+\omega)}\right)
     \label{eq25}
\end{eqnarray}

The equations presented above incorporate the equation of state parameter, denoted as $ p=\omega \rho $, which describes the relationship between pressure ($p$) and energy density ($\rho$) for different fluid components.

Here, the notation with a prime denotes differentiation with respect to $N$, which represents the logarithm of the scale factor $a$ (often referred to as the number of e-foldings).

To establish a connection between the model predictions and observations, it is possible to define several quantities of observational significance. One such quantity is the deceleration parameter $q$, which characterizes the rate of expansion of the Universe. The deceleration parameter can be expressed as follows:
\begin{eqnarray}
    q = -1 - \frac{\dot{H}}{H^2}= -\frac{3}{2} \frac{\beta (2n-1)\omega}{\alpha (n\omega - \omega -n)}y + \frac{3}{2 \alpha}x +\frac{1}{2}
    \label{eq26}
\end{eqnarray}

The effective equation of state (EoS) parameter $\omega_{eff}$ can be expressed as 

\begin{eqnarray}
    \omega_{eff} = \frac{1}{3}(2q-1)= -\frac{\beta (2n-1)\omega}{\alpha (n\omega - \omega -n)}y + \frac{x}{\alpha}
    \label{eq27}
\end{eqnarray}

 The values of $\omega$ determine the nature of the cosmic fluid, with $\omega = 1, \frac{1}{3}, 0, -\frac{1}{3}, \frac{-2}{3},$ and $-1$ corresponding to stiff matter, radiation, baryons, cosmic strings, domain walls, and a cosmological constant-like fluid, respectively. 

\subsection{Phase Space Analysis of model}
\label{sec4.1}
In order to analyze the system described by ordinary differential equations ($\ref{eq24}$) and ($\ref{eq25}$), our initial step involves identifying the critical points. These critical points correspond to the solution set of the ordinary differential equations, specifically when both $x'$ and $y'$ equal zero. To examine the behavior of the system in phase space, we investigate the stability of the critical points ($x_*, y_*$) by evaluating the eigenvalues of the Jacobian matrix at those points. For further details, please refer to \cite{W}.

The system of equations ($\ref{eq24}$) and ($\ref{eq25}$) is autonomous and possesses three critical points labeled as $A$, $B$, and $C$. Table  $\ref{table1}$ provides information about the cosmological parameters associated with these critical points ($A, B, C$), as well as the characteristics of the critical points and the eigenvalues of the Jacobian matrix at those points.

\begin{figure}[h!]
    \centering
    \includegraphics[width=4.0in ]{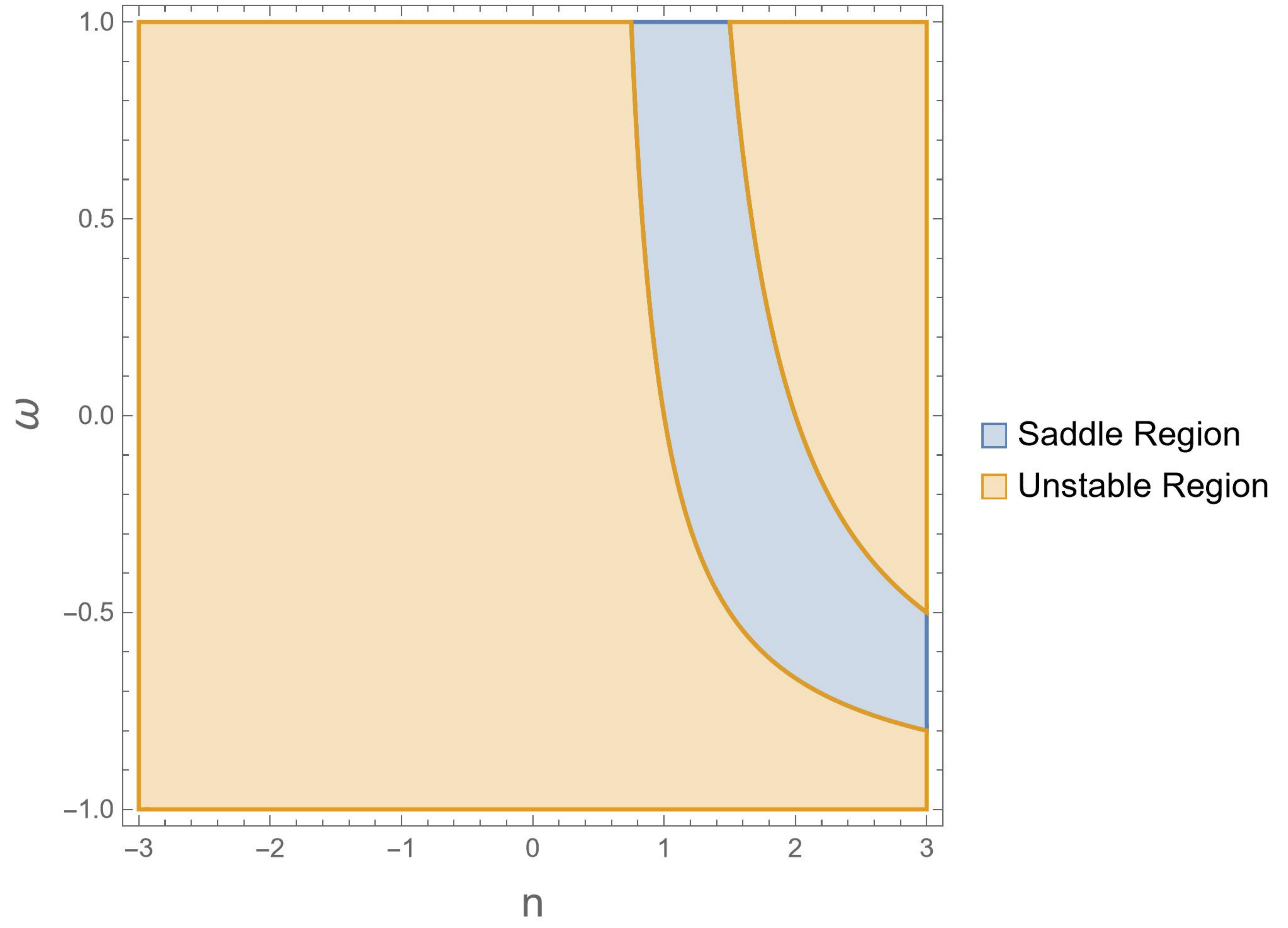}
    \caption{ Region plot which represents the nature of critical point `A' at different values of $n$ and $\omega$.}
    \label{figc1}
\end{figure}
\hfill

\subsection{Stability Analysis}
\label{sec4.2}
In this section, we will examine the stability of the model, which is characterized by the autonomous system of equations ($\ref{eq24}$) and ($\ref{eq25}$). This system encompasses three critical points, and we will now delve into a discussion of these specific critical points while considering the information presented in Table $\ref{table1}$.

Point A : The critical point $(0,0)$ is a constant presence in the model. The eigenvalues corresponding to this point are $\lambda_1 = 3$ and $\lambda_2 = 3 \left( \frac{1-(2n-1)(1+ \omega)}{1-(n-1)(1+ \omega)} \right) = \frac{3P}{Q}$, where $\omega$, $n$ are model parameters. It should be noted that the fraction $\frac{P}{Q}$ is negative when $\frac{1}{2} \left(1+\frac{1}{1+\omega} \right) < n < \left(1+\frac{1}{1+\omega} \right)$ for $\omega \neq -1$, and it is positive when $n$ lies in the range $\left(-\infty,\frac{1}{2} \left(1+\frac{1}{1+\omega}\right) \right) \cup \left(\left(1+\frac{1}{1+\omega}\right), \infty \right)$, with the exception of $\omega \neq -1$. As a result, this point will exhibit a saddle behavior when $\frac{1}{2} \left(1+\frac{1}{1+\omega} \right) < n < \left(1+\frac{1}{1+\omega} \right)$, and it will be unstable when $n$ falls within the range $\left(-\infty,\frac{1}{2} \left(1+\frac{1}{1+\omega}\right) \right) \cup \left(\left(1+\frac{1}{1+\omega}\right), \infty \right)$, provided that $\omega \neq -1$. Consequently, whenever $\omega$ equals $ -1$, this point will always act as unstable. The region plot for these cases is given in figure \ref{figc1}.

\begin{table}[h]
\centering
\begin{tabular}{|c|c|c|c|c|c|c|c|c|}
\hline
Point & $x_*$ & $y_*$ & $\lambda_1$ & $\lambda_2$ & $q$ & $\omega_{eff}$ & $r$ & $s$\\
\hline
\hline
A & $0$ & $0$ & $3$ & $\frac{3P}{Q}$ & $\frac{1}{2}$ & $0$ & $1$ & Undefined \\
\hline
B & $-\alpha$ & $0$ & $-3$ & $ \frac{-3n\left(1+\omega \right)}{Q}$ & $-1$ & $-1$ & $1$ & $0$ \\
\hline
C & $0$ & $\frac{\alpha (n\omega -\omega -n)P}{\beta (2n-1)\omega Q}$ & $\frac{3n\left(1+\omega \right)}{Q}$ & $- \frac{3 P}{Q}$ & $\frac{1}{2} \left(1- \frac{3P}{Q} \right)$ & $-\frac{P}{Q}$ & $r_1$ & $s_1$\\
\hline
\end{tabular}
\caption{Cosmological evolution and behavior of the model at critical points}

where  $r_1=-\frac{9 P n (\omega +1)}{2 Q^2}-\frac{9 P \left(\frac{P}{Q}-1\right)}{2 Q}+\left(2-\frac{3 P}{Q}\right) \left(\frac{1}{2}-\frac{3 P}{2 Q}\right)$ , $s_1= \frac{-2(r_1 - 1)Q}{9P}$, $P= 1-(2n-1)(1+ \omega)$ and $Q=1-(n-1)(1+ \omega)$.

\label{table1}
\end{table}

The corresponding deceleration parameter $q$ and $w_{eff}$ are $\frac{1}{2}$ and $0$ respectively. At this point, the Universe is undergoing the phase of decelerating expansion with scale factor $a \propto \left(\frac{3}{2} t - c_1 \right)^{\frac{2}{3}}$ where $c_1$ is an integration constant as $q > 0$ always. Additionally, the effective equation of the state parameter ($w_{eff}$) remains constantly equal to $0$, which represents a matter-dominated universe.

\begin{figure}[h!]
    \centering
    \includegraphics[width=4.0in ]{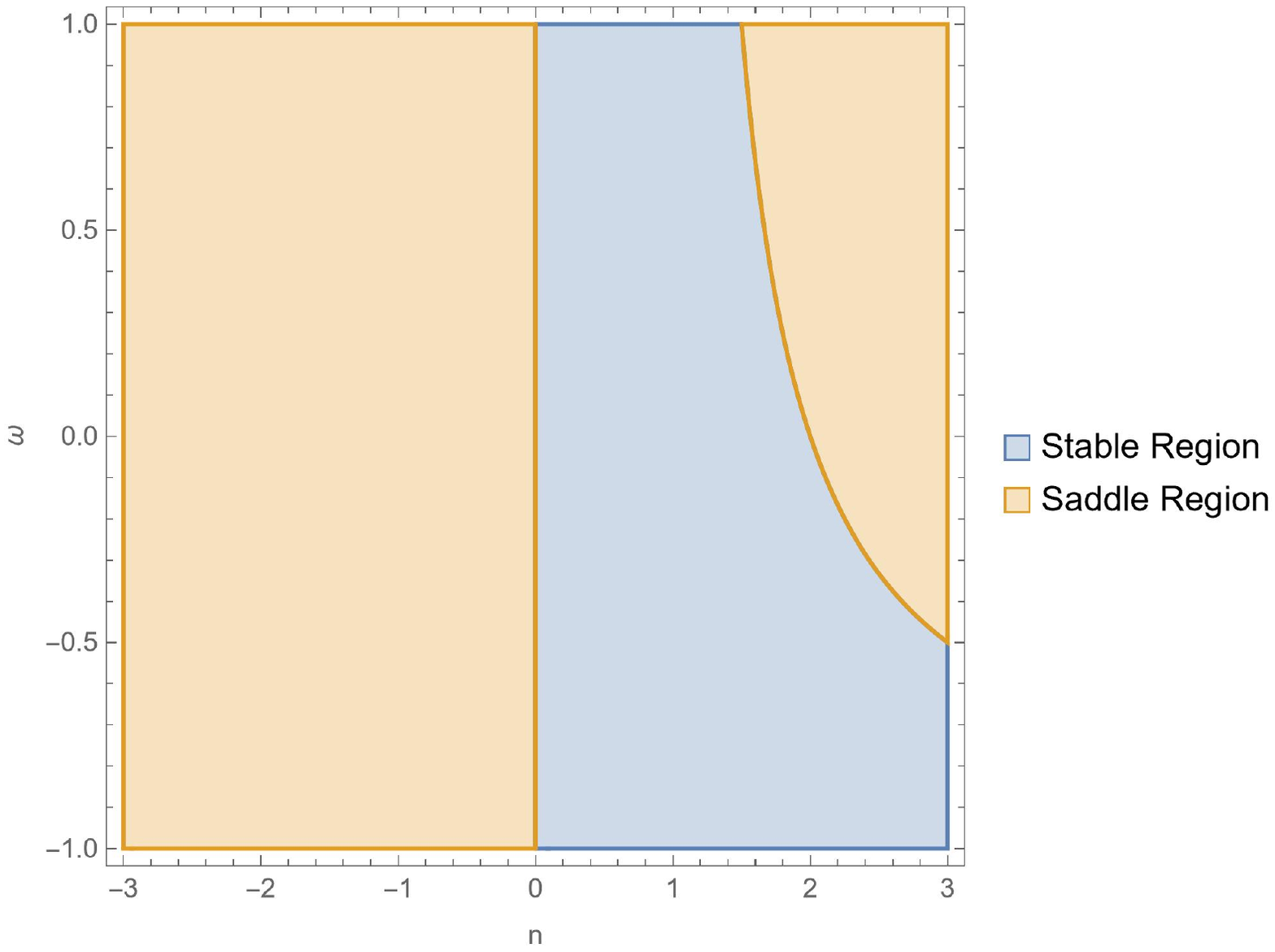}
    \caption{ Region plot which represents the nature of critical point `B' at different values of $n$ and $\omega$.}
    \label{figc2}
\end{figure}
\hfill

Point B : The model consistently includes the critical point $\left(-\alpha, 0 \right)$. The eigenvalues associated with this point are $\lambda_1 = -3$ and $\lambda_2 = \frac{-3n(1+\omega)}{1-(n-1)(1+\omega)}$, where $\omega$, $n$ are model parameters. This point is stable when $0<n<\left(\frac{2+\omega}{1+\omega} \right)$ and saddle otherwise provided $n \neq 0$ and $\omega \neq -1.$ We have visualise these cases in the region plot \ref{figc2}.

At the given point, the Universe is experiencing a phase of accelerating expansion with scale factor $a \propto e^{H_0 t}$ where $H_0$ is constant, characterized by the corresponding deceleration parameter $q$ having a value of $-1<0$. Consequently, the effective equation of the state parameter ($w_{\text{eff}}$) is always equal to $-1$. It signifies a specific scenario known as the cosmological constant or vacuum energy scenario. In this case, the Universe is dominated by an exotic form of energy called dark energy.

Point C : The critical point $\left(0, \frac{\alpha (n\omega -\omega -n)P}{\beta (2n-1)\omega Q} \right)$ is present as long as $(n-1)(1+\omega) \neq 1, \omega \neq 0$, $\beta \neq 0$, and $n \neq \frac{1}{2}$, where $\omega$, $n$ are model parameters. The point in question is characterized by eigenvalues, with $\lambda_1 = \frac{3n\left(1+\omega \right)}{Q} = \frac{3n(1+\omega)}{1-(n-1)(1+\omega)}$ and $\lambda_2 = -\frac{3 P}{Q}$, where $P = 1-(2n-1)(1+\omega)$ $\And$ $Q =1-(n-1)(1+\omega)$. This point will exhibit stability when the value of $n$ lies within the interval $\left(-\infty,0 \right) \cup \left(\frac{2+\omega}{1+\omega}, \infty \right)$. It will be unstable when $n$ belongs to the range $\left( \frac{2+\omega}{2+2\omega}, \frac{2+\omega}{1+\omega} \right)$, and it will act as a saddle for $n$ within the interval $\left(0, \frac{2+\omega}{2+2\omega} \right)$, provided that $\omega \neq -1$. The following region plot visualizes these cases based on the different values of $n$ and $\omega$:

\begin{figure}[h!]
    \centering
    \includegraphics[width=4.0in ]{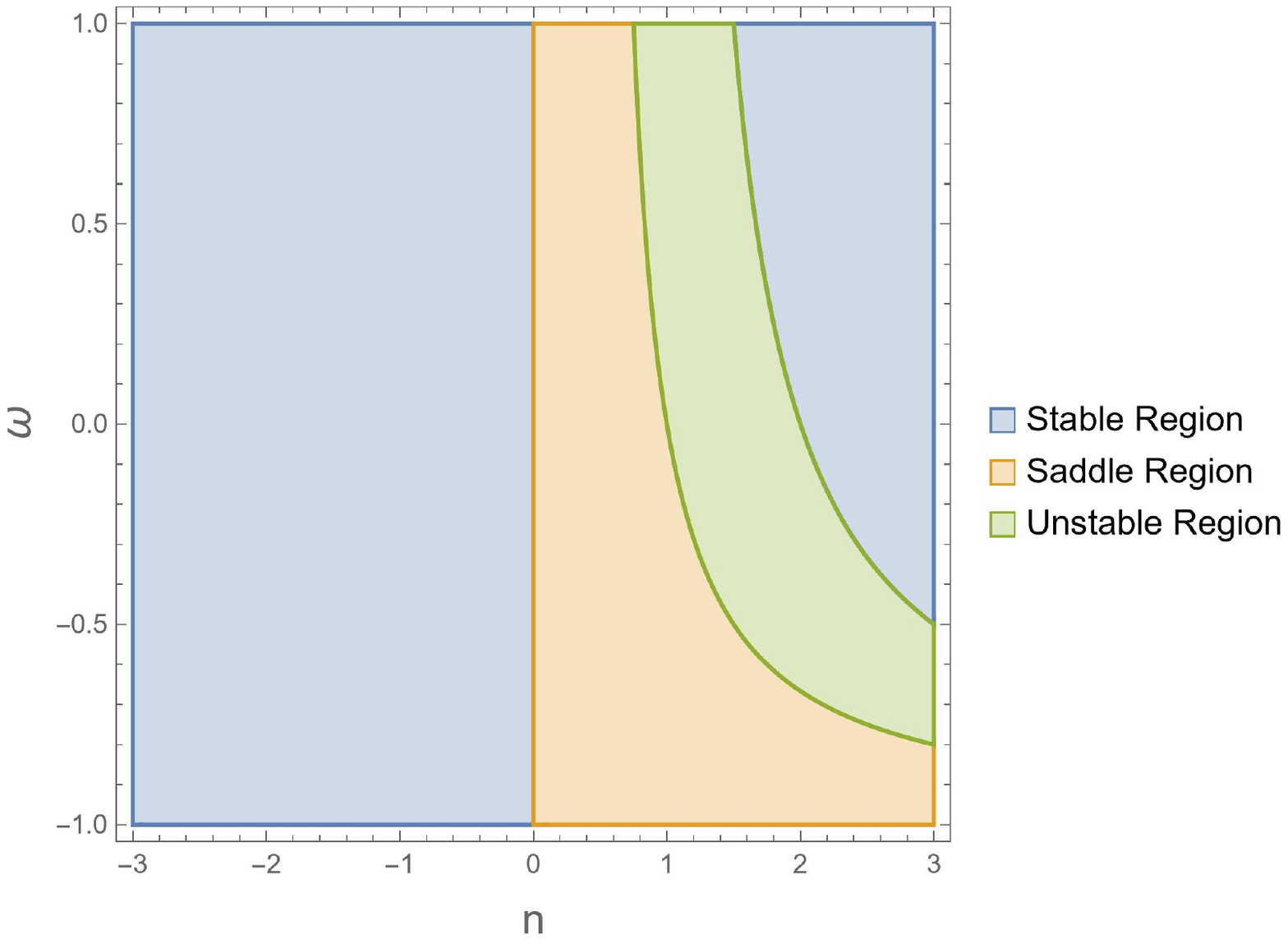}
    \caption{ Region plot which represents the nature of critical point `C' at different values of $n$ and $\omega$.}
    \label{figc3}
\end{figure}
\hfill

 In this context, the expansion of the Universe is characterized by the scale factor, $a \propto (kt + c_3)^{-\frac{1}{k}}$. Here, $k=\frac{3}{2} \left(\frac{P}{Q} - 1 \right) $ and $c_3$ represents an integration constant. The deceleration parameter $q$ can be expressed as $\frac{1}{2} \left(1- \frac{3P}{Q} \right)$ or equivalently as $\frac{1}{2} \left(\frac{(1+\omega)(5n-2) -2}{1-(n-1)(1+\omega)} \right)$. Consequently, the Universe will undergo a decelerating phase of expansion if the value of $n$ lies within the range $\left( \frac{4+2\omega}{5+5\omega}, \frac{2+\omega}{1+\omega} \right)$, provided $\omega \neq -1 $, and an accelerating phase otherwise. At this particular point, the effective equation of the state parameter can be represented as $w_{\text{eff}} = -\frac{P}{Q}$. As a result, the model is classified under the quintessence scenario when the value of $n$ falls within the range $\left(0,  \frac{4+2\omega}{5+5\omega} \right) $ and it falls into the phantom scenario when $n$ is within the range $\left(-\infty,0 \right) \cup \left(\frac{2+\omega}{1+\omega}, \infty \right)$, except the case when $\omega$ is not equal to $ -1$. The scenario where $\omega = -1$ corresponds to the cosmological constant like dark energy.

\begin{figure}[h!]
    \centering
    \includegraphics[width=3.0in]{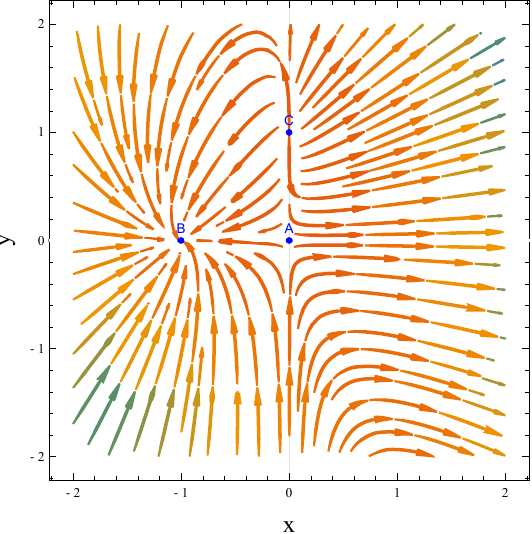}
    \caption{ Phase plane with the parameters $\alpha = 1$, $\beta = 1$, $n=1$, and $\omega = \frac{1}{3}$.}
    \label{fig1}
\end{figure}
\hfill

\begin{figure}[h!]
    \centering
    \includegraphics[width=3.0in]{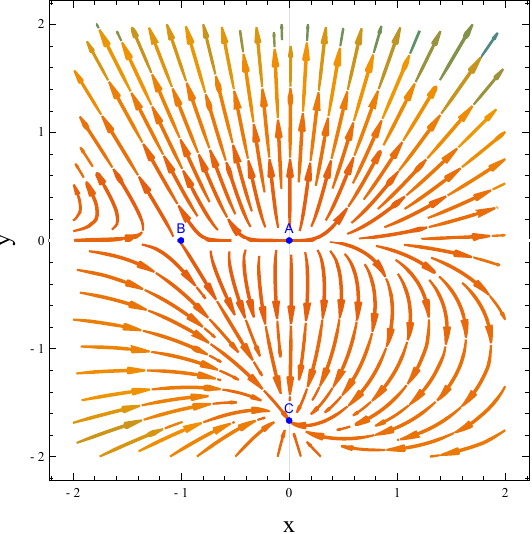}
    \caption{ Phase plane with the parameters $\alpha = 1$, $\beta = 1$,  $n=2$, and $\omega = 1$.}
    \label{fig2}
\end{figure}
\hfill

\begin{figure}[h!]
    \centering
    \includegraphics[width=3.0in]{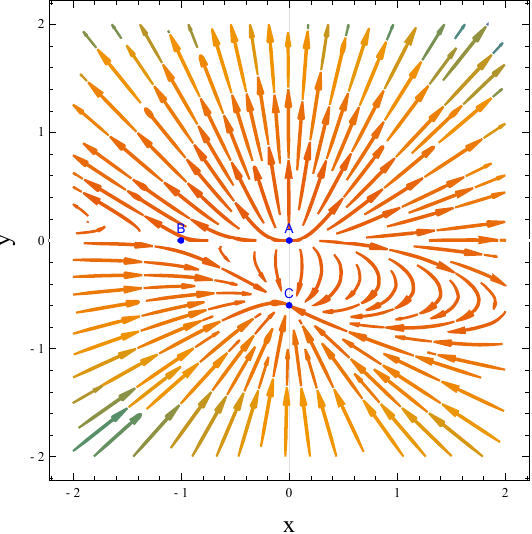}
    \caption{ Phase plane with the parameters $\alpha = 1$, $\beta = 1$, $n=3$, and $\omega = 1$.}
    \label{fig3}
\end{figure}
\hfill

\begin{figure}[h!]
    \centering
    \includegraphics[width=3.0in]{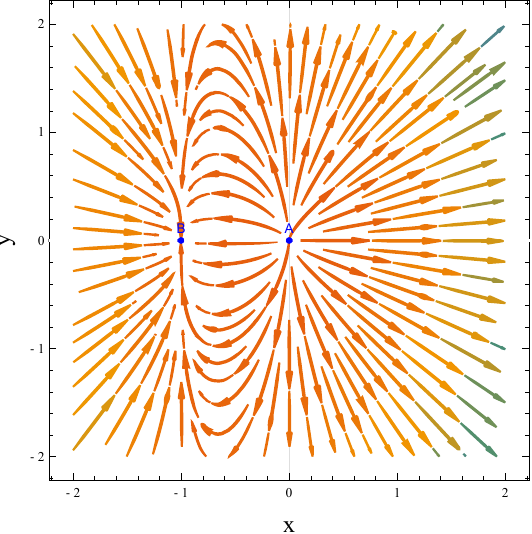}
    \caption{ Phase plane with the parameters $\alpha = 1$, $\beta = 1$,  $n=\frac{1}{2}$, and $\omega = 0$.}
    \label{fig4}
\end{figure}
\hfill

\begin{figure}[h!]
    \centering
    \includegraphics[width=3.0in]{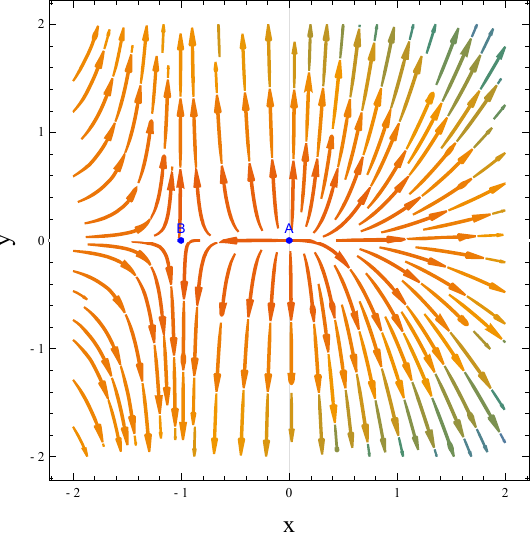}
    \caption{ Phase plane with the parameters $\alpha = 1$, $\beta = 1$,  $n=3$, and $\omega = 0$.}
    \label{fig5}
\end{figure}
\hfill
\begin{figure}[h!]
    \centering
    \includegraphics[width=3.0in]{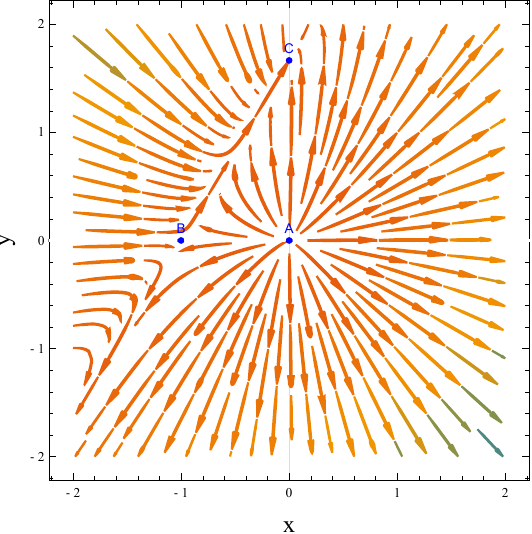}
    \caption{ Phase plane with the parameters $\alpha = 1$, $\beta = 1$,  $n=-1$, and $\omega = -\frac{1}{2}$.}
    \label{fig6}
\end{figure}
\hfill

\begin{figure}[h!]
    \centering
    \includegraphics[width=3.0in]{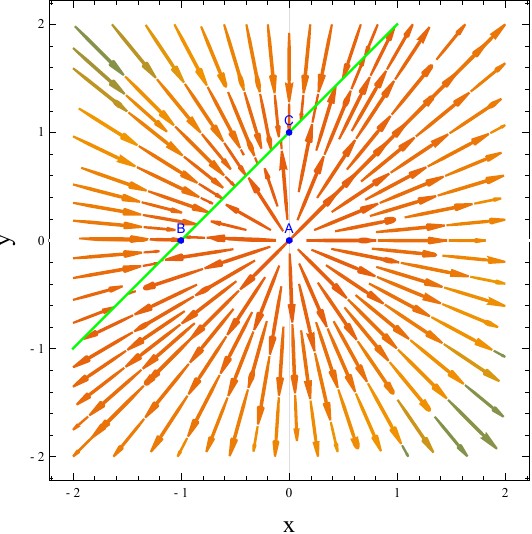}
    \caption{ Phase plane with the parameters $\alpha = 1$, $\beta = 1$,  $n=3$, and $\omega = -1$.}
    \label{fig7}
\end{figure}
\hfill

\clearpage


\subsection{Observational Constraints with Cosmic Chronometers (CC)}
The dataset used in our analysis is derived from the differential age method of $31$ data points in the redshift range $0.070 < z < 1.965$ \cite{Sharov+2018}. The cosmic chronometers approach is based on the equation $H(z)=\dot{a}/a=-(1+z)^{-1}dz/dt$, where, $a$ and $z$ are the scale factor and redshift, respectively. To find the maximum likelihood estimates of model parameters, we can define the likelihood function as $\mathcal{L}_{CC}\propto-\frac{1}{2}\chi_{_{CC}}^2$, such that \cite{sajal5}
\begin{equation}
    \chi_{_{CC}}^2=\sum_{i=1}^{31}\left( \frac{H_{th}(\theta,z_i)-H_{obs}(z_i) }{\sigma_{H(z_i)}}\right)^2
	\label{eqChi}
\end{equation}
where, $H_{th}(\theta,z_i)$ and $H_{obs}(z_i)$ represent the theoretical and observed values of the Hubble parameter at redshift $z_i$, respectively, and $\sigma_{H(z_i)}$ denotes the associated error in the observed Hubble parameter values. Using cosmic chronometer relation, we can rewrite the equations (\ref{eq19}) as
\begin{eqnarray}
    \frac{\rho'}{\rho} - 3 (1+ \omega)\frac{1}{1+z} = (n-1)(1+\omega)\frac{\rho'}{\rho}
    \label{eq32}
\end{eqnarray}
Here, the prime ($'$) represents differentiation with respect to redshift $z$, i.e., $\rho' = \frac{d \rho}{dz}$.
After performing the integration of equation (\ref{eq32}), we obtain the solution as
\begin{eqnarray}
    \rho = \rho_0 (1+z)^{\frac{3(1+\omega)}{2-n+\omega- n \omega}}
    \label{eq33}
\end{eqnarray}
 where $\rho_0$ is an integrating constant. We can rewrite the equation (\ref{eq20}) as
\begin{eqnarray}
\frac{\Lambda}{3 {H_0}^2} \frac{(-\omega)^{1-n}}{(n \omega -n -\omega)} + \beta \frac{\rho^n}{3 {H_0}^2} = \frac{\alpha (-\omega)^{1-n}}{(n + \omega -n \omega)} \frac{H^2}{{H_0}^2}
\label{eq34}
\end{eqnarray}
where, $H_0$ is the current value of the Hubble parameter at $z=0$. We define the critical densities for dark matter and dark 
 energy as $\Omega_{m_0}=\frac{\rho^n_0}{3 {H_0}^2}$ and $\Omega_{\Lambda_0}=\frac{\Lambda}{3 {H_0}^2}$, respectively. Using these relations and equation (\ref{eq33}), we can rewrite the equation (\ref{eq34}) as
\begin{eqnarray}
    -\Omega_{\Lambda_0} \left( \frac{(-\omega)^{1-n}}{n + \omega - n \omega}\right) + \beta \Omega_{m_0} (1+z)^{\frac{3 n (1+\omega)}{2-n+\omega- n \omega}} = \alpha\frac{H^2}{{H_0}^2} \left( \frac{(-\omega)^{1-n}}{n + \omega - n \omega}\right)
    \label{eq35}
\end{eqnarray}
At $z=0$, equation (\ref{eq35}) becomes
\begin{equation}
    -\Omega_{\Lambda_0}\left( \frac{(-\omega)^{1-n}}{n + \omega - n \omega}\right) + \beta \Omega_{m_0} = \alpha \left( \frac{(-\omega)^{1-n}}{n + \omega - n \omega}\right)
    \label{eq37a}
\end{equation}
Using equations (\ref{eq35}) and (\ref{eq37a}), we may write the Hubble parameter in terms of redshift as 
\begin{eqnarray}
H(z) = H_0 \sqrt{\left(\frac{\beta}{\alpha} \right) \Omega_{m_0}(n + \omega -n \omega) (- \omega)^{n-1} \left[(1+z)^{\frac{3 n (1+\omega)}{2-n+\omega- n \omega}} -1\right] + 1}
    \label{eq37}
\end{eqnarray}
 
\begin{figure}[h!]
    \centering
    \includegraphics[width=4.0in ]{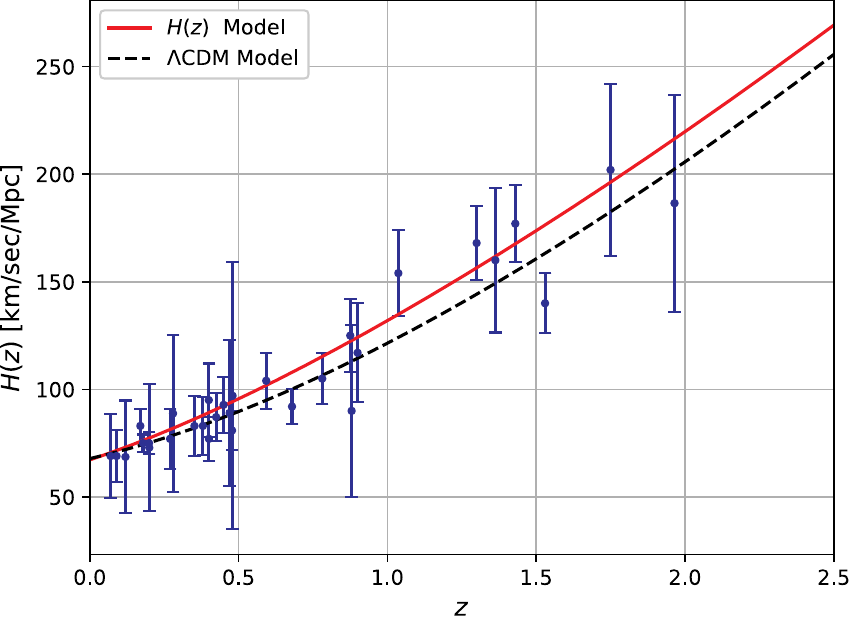}
    \caption{Hubble parameter $H(z)$ as a function of redshift $z$, plotted using 31 observational Hubble data points with error bars. The solid curve represents the best fit $H(z)$ curve for the central values of the parameters $H_0 = 67.2~ km/sec/Mpc, n = -1.24, \alpha = -0.059, \beta = 0.05, \Omega_{m_0} = 0.25$, and $\omega = -1.278$ obtained from MCMC analysis.}
    \label{figHZ}
\end{figure}

\begin{figure}[h!]
    \centering
    \includegraphics[width=5.1in ]{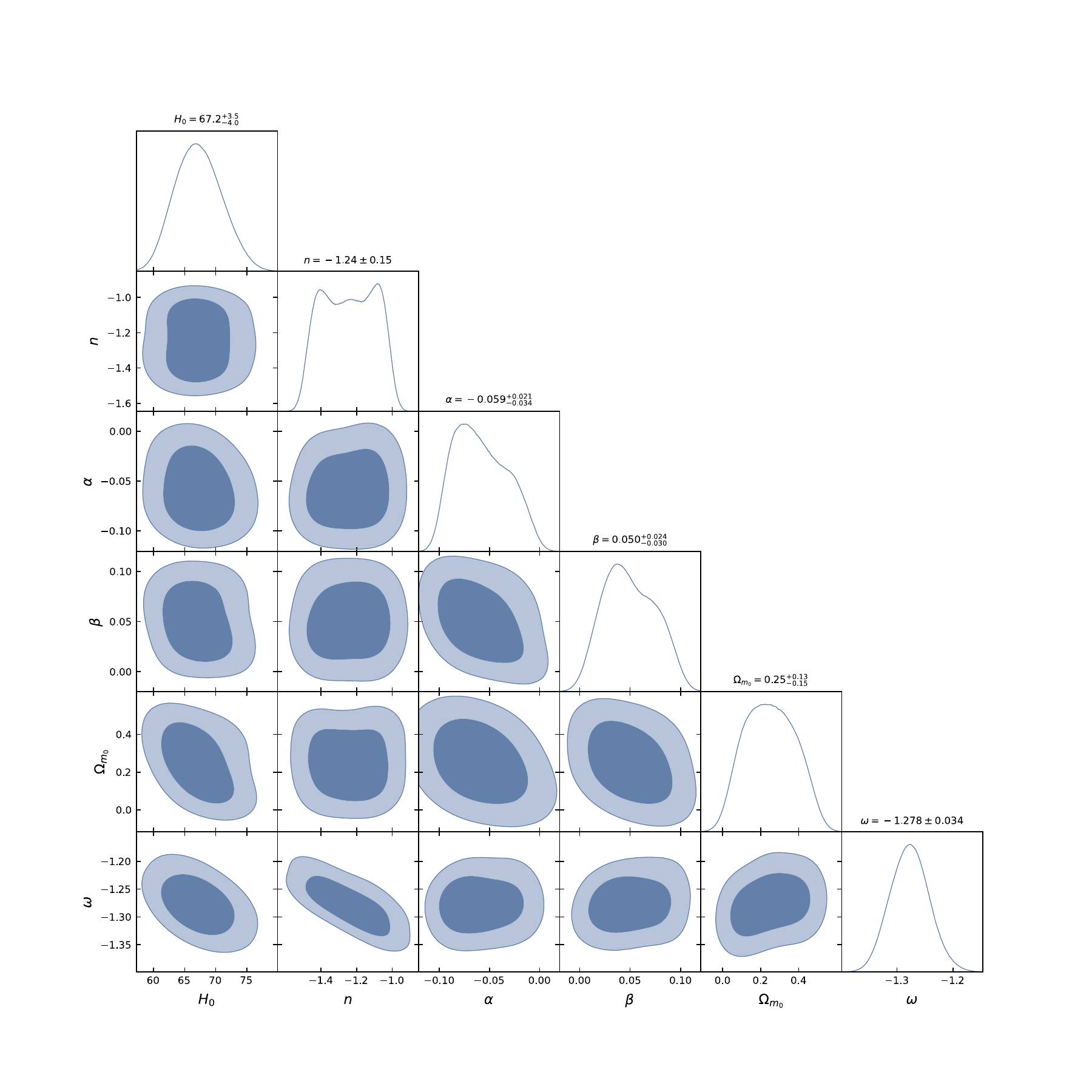}
    \caption{Marginalized $1\sigma$, $2\sigma$ contours with $1$ dimensional posterior distributions for parameters $H_0, n, \alpha,  \beta,  \Omega_{m_0} $, and $\omega$.}
    \label{figContour}
\end{figure}

To constrain the set of model parameters $\lbrace H_0, n, \alpha, \beta, \Omega_{m_0}, \omega \rbrace$ subjected to the cosmic chronometer data we use Markov Chain Monte Carlo (MCMC) sampling method in emcee python tool \cite{Foreman+2013}.  The analysis incorporates priors as
	$ H_0 \in (60, 80), n \in (-1.5, -1.0), \alpha \in (-0.1, 0),  \beta \in (0, 0.1), \Omega_{m_0} \in (0, 0.5)$,  and $\omega \in (-2, 2)$. The median values of posterior probability distributions within $1\sigma$ confidence level are obtained as $ H_0=67.2^{+3.5}_{-4.0}, n=-1.24\pm0.15, \alpha=-0.059^{+0.021}_{-0.034},  \beta=0.050^{+0.024}_{-0.030}, \Omega_{m_0}=0.25^{+0.13}_{-0.15}$,  and $\omega=-1.278\pm0.034$, and the minimum value of $\chi_{_{CC}}^2$ is $16.49$. The best fit $H(z)$ curve of 
 the model is compared with the $\Lambda$CDM model in Figure \ref{figHZ} for the central values of the model parameters obtained using MCMC analysis. The evolution of dark energy in both models are in different ways. The constrained value of $H(z) = 67.2\  \text{km/s/Mpc}$ is close to $\Lambda$CDM model estimate  $H_0 = 67.36 \pm 0.54\ \text{km/s/Mpc}$ from Planck results \cite{NA}. The $1\sigma$ and $2\sigma$ contour plots for the model parameters are portrayed in Fig. \ref{figContour}.

\subsection{Statefinder Diagnostic}
\label{sec4.3} 
In this work, we present the statefinder diagnostic in two distinct approaches: one approach uses the dynamical variables, and the other is based on cosmological redshift. We analyze the resulting cosmological quantities by applying the constrained parameter values obtained from observational data.

The statefinder diagnostic tools consist of geometric parameters that enable the investigation of dark energy properties in a model-independent manner. The statefinder parameters are precisely defined as \cite{SSS}
\begin{eqnarray}
    r= q+2q^2- \frac{\dot{q}}{H} , \hspace{0.3cm}  s= \frac{r-1}{3(q-\frac{1}{2})}
    \label{eq28}
\end{eqnarray}
Dimensionless parameters $\{r,s\}$ are derived directly from the scale factor and its derivatives. When $r = 1$ and $s = 0$, it corresponds to the $\Lambda$CDM model, while $r = 1$ and $s = 1$ represent the Standard Cold Dark Matter (SCDM) model. However, for evolving dark energy models, the $r$ value does not equal $1$.

The trajectories of the Chaplygin gas model and quintessence model lie in different regions of the $r-s$ plane. Specifically, the Chaplygin gas model trajectories can be found in the region where $r > 1$ and $s < 0$, while the quintessence model trajectories are located in the region where $r < 1$ and $s > 0$.

Using dynamical variable, the statefinder parameters $r$ and $s$ can be expressed as:

\begin{eqnarray}
\begin{split}
    r=\left(\frac{-3}{2} \frac{\beta (2n-1)\omega }{\alpha (n\omega -n -\omega) }y + \frac{3}{2 \alpha} x + \frac{1}{2}\right) \left(2 -\frac{3 \beta (2n-1)\omega }{\alpha (n\omega -n -\omega) }y + \frac{3}{\alpha} x  \right) \\
    - \frac{9}{2}\left(\frac{\beta (2n-1)\omega }{\alpha (n\omega -n -\omega) }y- \frac{x}{\alpha} -1 \right) \left(\frac{\beta (2n-1)\omega }{\alpha (n\omega -n -\omega) }y - \frac{x}{\alpha} \right) \\- \frac{9}{2} \frac{\beta n (2n-1) \omega (1+\omega) }{\alpha (n\omega -n -\omega)(1-(n-1)(1+\omega)) }y
    \label{eq29}
\end{split}
\end{eqnarray}

\begin{eqnarray}
    s= \frac{-2}{9} \left(\frac{r-1}{\frac{\beta (2n-1)\omega }{\alpha (n\omega -n -\omega) }y - \frac{x}{\alpha}} \right)
    \label{eq30}
\end{eqnarray}
In the present model, varying the model parameters can lead to diverse characteristics of interacting dark energy. Exploring the late-time attracting behavior of the model associated with critical points $A$, $B$, and $C$ would be a fascinating area of investigation.

\begin{figure}[h]
    \centering
    \includegraphics[width=3.5in ]{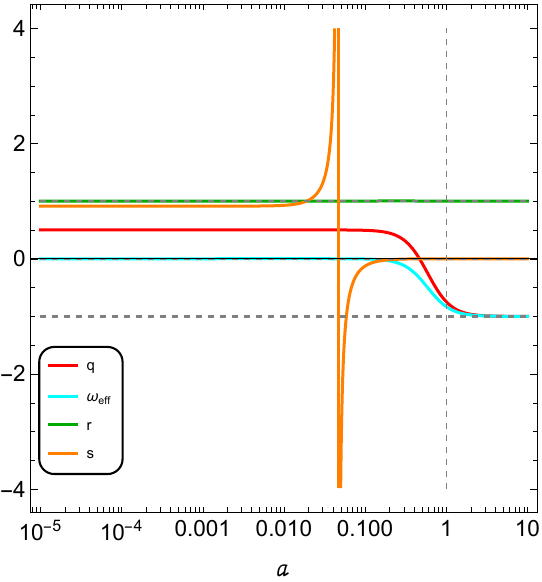}
    \caption{Evolution of cosmological quantities $q$, $\omega_{eff}$, $r$, and $s$ with scale factor $a$ for the central values of the parameters $n = -1.24, \alpha = -0.059, \beta = 0.05$, and $\omega = -1.278$ obtained from MCMC analysis.}
    \label{fig9}
\end{figure}

At critical point $A$, the value of $r$ is $1$, while the value of $s$ remains unspecified. On the other hand, at critical point B, the parameters $r$ and $s$ are fixed at $1$ and $0$, respectively, indicating that the model consistently adheres to the $\Lambda$CDM paradigm. At critical point $C$, the values of $r$ and $s$ can be calculated as follows:

$r=-\frac{9 P n (\omega +1)}{2 Q^2}-\frac{9 P \left(\frac{P}{Q}-1\right)}{2 Q}+\left(2-\frac{3 P}{Q}\right) \left(\frac{1}{2}-\frac{3 P}{2 Q}\right)$ , $s= \frac{-2(r - 1)Q}{9P}$, where $P = 1-(2n-1)(1+ \omega)$ and $Q =1-(n-1)(1+ \omega)$. 
Interestingly, when the parameter values are set to $\omega = \frac{1}{2}$ and $n = 0$, the values of $r$ and $s$ become $1$ and $0$, respectively. This implies that the model aligns with the  $\Lambda$CDM paradigm at these specific parameter values. To portray the evolution of cosmological quantities, we employ the numerical solution techniques in Figure $\ref{fig9}$.

In terms of redshift, the statefinder diagnostic  parameters $\{r,s\}$, defined in equation (\ref{eq28}), can be expressed as follows \cite{sajal3}:
\begin{eqnarray}
    r = 1 - 2(1+z) \frac{H'}{H} + (1+z)^2 \frac{H''}{H} + (1+z)^2 \left(\frac{H'}{H}\right)^2 \\
    s = \frac{-2(1+z)\frac{H'}{H} + (1+z)^2\frac{H''}{H} + (1+z)^2\left(\frac{H'}{H}\right)^2}{3\left((1+z)\frac{H'}{H} - \frac{3}{2}\right)}
\label{eq39}
\end{eqnarray}

\begin{figure}[h!]
    \centering
    \includegraphics[width=4.0in ]{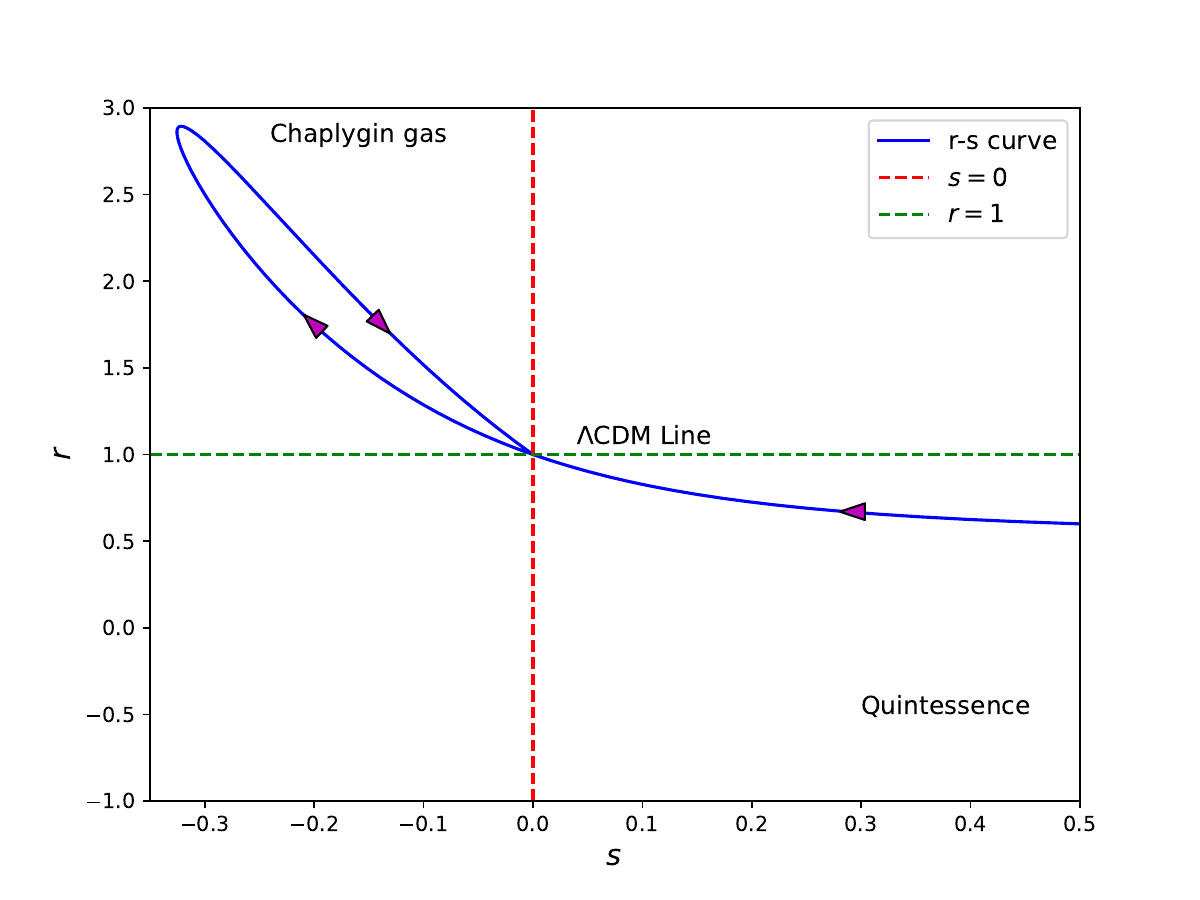}
    \caption{Trajectory in the {$r - s$}
 plane for the central values of the parameters in MCMC analysis.}
    \label{figrsbestfit1}
\end{figure}

We plot the universe's trajectory in the $r - s$ plane for the range $-1.0 \leq z \leq 0.6$ in Figure \ref{figrsbestfit1}. This figure shows that our model exhibits quintessence-like properties in the recent past, characterized by $r < 1$ and $s > 0$. It transitions to the $\Lambda$CDM model at a certain point and then follows the Chaplygin gas model, characterized by $r > 1$ and $s < 0$.

\newpage

\section{Conclusion}
\label{sec5}
In this study, we have explored the cosmic expansion of the Universe at later stages by examining the modified $f(R, L_m)$ gravity theory. Our investigation focused on a non-linear $f(R, L_m)$ model, denoted as $f(R, L_m) = \Lambda + \frac{\alpha}{2} R + \beta L_m^n $, where the parameters $n$, $\Lambda$, $\alpha$ and $\beta$ are unrestricted within the model. We have explored an autonomous system described by equations ($\ref{eq24}$) and ($\ref{eq25}$) in our cosmological $f(R, L_m)$ model. After conducting a dynamical systems analysis with the equation of state $p = \omega \rho$, we have determined and presented the eigenvalues, critical points in Table 1, and their existence for this model are given in Figures ($\ref{figc1}$), ($\ref{figc2}$) and ($\ref{figc3}$). It is worth to notice here that the critical point A is same for our model and model proposed by Harko \& Lobo \cite{H4}, whereas critical points B \& C are dependent on parameters $\alpha, \beta$ and $n$. It is noticed that the eigenvalues, deceleration, equation of state, $\{r,s\}$ parameters are dependent on free parameters ($\alpha, \beta$ and $n$) and tend to the model proposed by Harko \& Lobo \cite{H4} when $\alpha=\beta=n=1$. To provide a clearer visualization of the universe's evolution, we generated stream plots for various parameter values of $n$ and $\omega$, as depicted in Figures $\ref{fig1}$, $\ref{fig2}$, $\ref{fig3}$, $\ref{fig4}$, $\ref{fig5}$, $\ref{fig6}$, and $\ref{fig7}$.  We also note that point $A$ represents radiation dominated era when $n=-1,\frac{1}{2}, 1, 2, 3$ and $\omega = 0, -\frac{1}{2}, -1, \frac{1}{3}$ whereas points $B$ and $C$ represent dark energy dominated universe when $n=-1,\frac{1}{2}, 1, 3$, $\omega = 0, -\frac{1}{2}, -1$, and $n= 2, 3$, $\omega = -1, \frac{1}{3}$ respectively. In addition to the analysis of the phase space, we further explore the evolution of the state-finder parameters to determine wether the present model aligns with $\Lambda$CDM model or not. The observation from Table 1 for the parameter $q$ for model give the information about the expansion of the Universe. We have observed the early decelerating stage of the Universe (at point A) and the transition phase (at point C) to accelerated stage of the Universe (at point B). It is noted that points B and C  behaves as late time attractor for suitable choices of parameter values of $n$ and $\omega$. It is also confirmed that the value of deceleration parameter $q=-1$ (de Sitter Universe) and  $\{r,s\}$ parameter is $\{1,0\}$ ($\Lambda$CDM) at critical point B. For constrained parameter values of $n$ and $\omega$, the Figure \ref{figrsbestfit1} shows that the universe's trajectory in the $r - s$ plane indicates that our model exhibits quintessence-like properties in the recent past then transitions to the $\Lambda$CDM model at a certain point and then follows the Chaplygin gas model. We have obtained the Hubble parameter in terms of redshift ($z$) and employed emcee Python package to portray the posterior distribution of the parameters using cosmic chronometer data.
\par
The observational constraints obtained in the present model suggest that the model behaves differently as compared to the $\Lambda$CDM model. In $\Lambda$CDM model, $ \omega = -1$, while in the present model $\omega$ has been constrained as $\omega = -1.278 \pm 0.034$. In other words, the present $f(R, L_m)$ model, governed by $H(z)$ (\ref{eq37}), predicts universe evolution under the influence of phantom dark energy. As a consequence, the energy density of the Universe will increase in the model, and the Universe may have a finite-time future singularity of big-rip type \cite{Caldwell2002, Nojiri2005, Haro2023}. In summary, the dark energy yielded by the `geometrical sector' in the present $f(R, L_m)$ model drives the universe's expansion in such a way that the EoS parameter of this energy is less than $-1$. The EoS parameter is different from the standard cosmological model having dark energy equation of state parameter $-1$. The value of $\omega$ subjected to the cosmic chronometer data suggests that the model may have a different evolution history in the future as compared to the standard cosmological model ($\Lambda$CDM model). It would be interesting to test the compatibility of model by also incorporating radiation as cosmic fluid with a wider set of observational data. We aim to study these issues in some near future.  

 \section*{Acknowledgments}
   The authors express their sincere thanks to the referees for valuable comments and
suggestions to improve the quality of the manuscript. AS expresses his sincere thanks to C.S.I.R., New Delhi, for providing financial support under CSIR (JRF) scheme through award No. 09/0013(18879)/2024-EMR-I. RC thanks SERB, New Delhi, for financial assistance through project No. CRG/2023/004560 (P-07/1328).\\
\\
Data Availability: The paper has no associated data. All concepts and logical implications are given in the
manuscript.

\end{document}